\documentclass[prb,amsmath,amssymb,floatfix]{revtex4}
\usepackage{graphicx}% Include figure files
\usepackage{bm}% bold math

\usepackage{amssymb,amsfonts,amsmath}

\def \>{\rangle} 
\def \<{\langle} 
 
\def\be{\begin{equation}} 
\def\ee{\end{equation}} 
\def\longrightharpoonup{\relbar\joinrel\rightharpoonup}
\def\longleftharpoondown{\leftharpoondown\joinrel\relbar}

\def\longrightleftharpoons{
  \mathop{
    \vcenter{
      \hbox{
      \ooalign{
        \raise1pt\hbox{$\longrightharpoonup\joinrel$}\crcr
	  \lower1pt\hbox{$\longleftharpoondown\joinrel$}
	  }
      }
    }
  }
}

\newcommand \bea {\begin{eqnarray}} 
\newcommand \eea {\end{eqnarray}}

\begin{document}

\title{Efficiency bounds for nonequilibrium heat engines}

\author{Pankaj Mehta and  Anatoli Polkovnikov}

\affiliation{Department of Physics, Boston University, 590 Commonwealth Ave., Boston, MA 02215\\
e-mail:asp@bu.edu; tel: 617-3584394;  fax: 617-353-9393}

\begin{abstract} 

We analyze the efficiency of thermal engines (either quantum or classical) working with a single heat reservoir like atmosphere. The engine first gets an energy intake, which can be done in arbitrary non-equilibrium way e.g. combustion of fuel. Then the engine performs the work and returns to the initial state. We distinguish two general classes of engines where the working body first equilibrates within itself and then performs the work (ergodic engine) or when it performs the work before equilibrating (non-ergodic engine). We show that in both cases the second law of thermodynamics limits their efficiency. For ergodic engines we find a rigorous upper bound for the efficiency, which is strictly smaller than the equivalent Carnot efficiency. I.e. the Carnot efficiency can be never achieved in single reservoir heat engines. For non-ergodic engines the efficiency can be higher and can exceed the equilibrium Carnot bound. By extending the fundamental thermodynamic relation to nonequilibrium processes, we find a rigorous thermodynamic bound for the efficiency of both ergodic and non-ergodic engines and show that it is given by the relative entropy of the non-equilibrium and initial equilibrium distributions.These results suggest a new general strategy for designing more efficient engines. We illustrate our ideas by using simple examples.

\end{abstract}

\maketitle

%%%%%%%%%%%%%%%%%%%%%%%%%%%%%%%%%%%%%%%%%%%%%%%%%%%%%%%%%%%%%%%%

Heat engines are systems that convert heat or thermal energy into macroscopic work.  Heat engines  play a major role in modern technology and are crucial to our understanding of thermodynamics \cite{reif1965fundamentals,fermi1936thermodynamics}.  Examples of heat engines include conventional combustion engine such as those found in  cars and airplanes, various light emitting devices, as well as naturally occurring engines such as molecular motors  \cite{magnasco1993forced}.  A conventional heat engine consists of two heat reservoirs, a hot reservoir that serves as a  source of energy and a cold reservoir that serves as an entropy sink.  The efficiency of  such engines is fundamentally limited by the second law of thermodynamics providing an upper bound given by the efficiency of a Carnot engine operating at the same temperatures \cite{reif1965fundamentals},
\be
\eta_c=1-{T_c \over T_h},
\label{carnot_engine}
\ee 
with $T_c$ and $T_h$ the temperature of the cold and hot reservoirs respectively. 

Real engines often differ significantly from the idealized, two-reservoir engines considered in classical thermodynamics. They operate with a single bath, such as the atmosphere,  that serves as an entropy sink.  Instead of a high temperature bath, energy is suddenly deposited in the system at the beginning of each cycle and is converted into mechanical work. The most common example of this are combustion engines such as those found in cars where energy is deposit in the system through the combustion of a fuel. Currently, the most realistic models describing combustion engines are based on the Otto cycle~\cite{reif1965fundamentals}, with a corresponding efficiency. which is less than  $\eta_c$ with appropriately chosen temperatures $T_c$ and $T_h$. One can ask some natural questions:  is the Carnot efficiency  a good  bound for the efficiencies of such single-reservoir engines or are these engines better described by a different bound? Are there realistic processes that allow you to realize these bounds? Can we overcome the thermodynamic bounds if we use engines which are not completely ergodic?

To address these questions, we generalize the fundamental relations of thermodynamics to describe large, nonequilibrium quenches in systems coupled to a thermal bath. We use these relations to derive new bounds for the efficiency of nonequilibrium engines that operate with a single bath. We analyze our bounds in two different regimes, a local equilibrium regime where the system quickly thermalizes with itself (but not the bath), and a non-erdgodic regime where the thermalization times are much longer than time scales on which work is performed. We demonstrate our results using simple examples such as an ideal gas that drives a piston and a magnetic gas engine.

The paper is organized as follows. In Sec.~\ref{sec:identities} we formulate generalized thermodynamic identities, which extend the fundamental thermodynamic relations to arbitrary non-equilibrium processes and introduce the notion of the relative entropy (or Kullback-Leibler divergence). In Sec.~\ref{sec:effic} we apply these results for finding the maximum efficiency of the non-equilibrium engines. We separately discuss bounds for ergodic (equilibrium) engines, non-ergodic (nonequilibrium) incoherent engines and non-ergodic coherent engines. Then in Sec.~\ref{sec:examples} we illustrate our results using simple examples and show that non-ergodic engines can indeed have higher efficiency than the ergodic ones. In Sec.~\ref{sec:derivation} we give rigorous derivation of the thermodynamic identities of the paper. Then In Sec.~\ref{sec:derivation_effic} we give the details of the derivation of the efficiencies of the ergodic and non-ergodic engines.

\section{Generalized Thermodynamic Identities} 
\label{sec:identities}

Most applications of thermodynamics are connected to the fundamental thermodynamic relation~\cite{kardar2007statistical}
\be
dE=TdS-\mathcal F d\lambda,
\label{fund_rel}
\ee
where $E$ is the energy if the system, $T$ is the temperature, $S$ is the entropy, $\lambda$ is some external macroscopic parameter, and $\mathcal F$ is the generalized force. When $\lambda$ is the volume $\mathcal F$ stands for the pressure and the fundamental relation takes the most familiar form $dE=TdS-PdV$. The fundamental relation mathematically encodes the fact that the energy of a system in equilibrium is a unique function of the entropy and external parameters. For quasistatic processes, one can associate the first term with the supplied heat and the second term with the work done on the system by changing the parameter $\lambda$. The fundamental relation can also be integrated for quasistatic processes and one can explicitly compute the total work, heat etc. However, how to generalize these calculations to strongly nonequilibrium processes where changes in energy, entropy, etc. can be large, is  still largely an open question.  Using the second law of thermodynamics, one can prove various inequalities. In particular, if we prepare a system $A$ in a thermal state with temperature $T_A$ and let it equilibrate with a bath at temperature $T$ then the second law of thermodynamics implies two related inequalties: (\label(see Sec.~\ref{sec:derivation})
\be
 T_A\Delta S_A-\Delta E_A\leq 0,\quad T\Delta S_A-\Delta E_A\geq 0.
\label{fund_rel1}
\ee
 The first inequality is also applicable to the case where an energy $\Delta E_A$ is deposited in the system in a nonequilibrium fashion, for example, by  an external energy pulse (then $T_A$ is the initial temperature of the system $A$), and the second inequality describes the relaxation of a system back to equilibrium. It implies that the free energy of the system can only go down during the relaxation~\cite{kardar2007statistical}. 

In this work we establish two main closely related results, which refine the inequalities (\ref{fund_rel1}) to arbitrary non- equilibrium protocols using the concept of relative entropy . Relative entropy, or the Kullback-Leibler Divergence, is well known in information theory \cite{cover1991elements, mackay2003information} and appears naturally in statistical mechanics within the context of large deviation theory \cite{touchette2009large}. In deriving our results, we will use ``quantum'' notations and restrict ourselves to discrete probability distributions.  Our results also equally apply  to classical systems with continuous probability distributions and can be derived from the corresponding ``quantum'' results by multiplying all distributions by an appropriately chosen density of states (see Ref.~[\onlinecite{bunin2011universal}] and Sec.~\ref{sec:derivation}). These general results valid for both quantum and classical systems are closely related to those recently obtained by S. Deffner and E. Lutz~\cite{deffner2010generalized,deffnerlutz2011} for quantum systems but deviate in a way that is crucial to our discussion (see Sec.~\ref{sec:derivation} for details). 

Consider a system $A$ with external parameter $\lambda=\lambda^{(1)}$ and a $\lambda$-dependent energy spectrum $\mathcal E_n^{(1)}\equiv \mathcal E_n(\lambda^{(1)})$ which is coupled to a thermal bath at temperature $T$. We assume that the bath is insensitive to the parameter $\lambda$ (Figure 1). Initially, the system is prepared in equilibrium with the bath and is described by a Boltzmann distribution of the form
\[
p_n^{(1)}= \exp[-\mathcal E_n^{(1)}/T]/Z_1,\; {\rm with},\; Z_1= \sum_n e^{-\beta \mathcal E_n^{(1)}}.
\] 
In stage $I$, the system undergoes an arbitrary process where $\lambda$ is changed from $\lambda^{(1)}$ to $\lambda^{(2)}$ , resulting in a new non-equilibrium state, characetrized by some generally non-equilibrium probability occupations of the energy eigenstates $q_n$. We do not assume that during this process the system is thermally isolated. Then in stage $II$, the system re-equilibrates with the bath, eventually reaching a new Boltzmann distribution with $\lambda=\lambda^{(2)}$, 
\[
p_n^{(2)}=exp[-\mathcal E_n^{(2)}/T]/Z_2,\quad Z_2= \sum_n e^{-\beta \mathcal E_n^{(2)}}.
\]
\begin{figure}
\includegraphics[width=15cm]{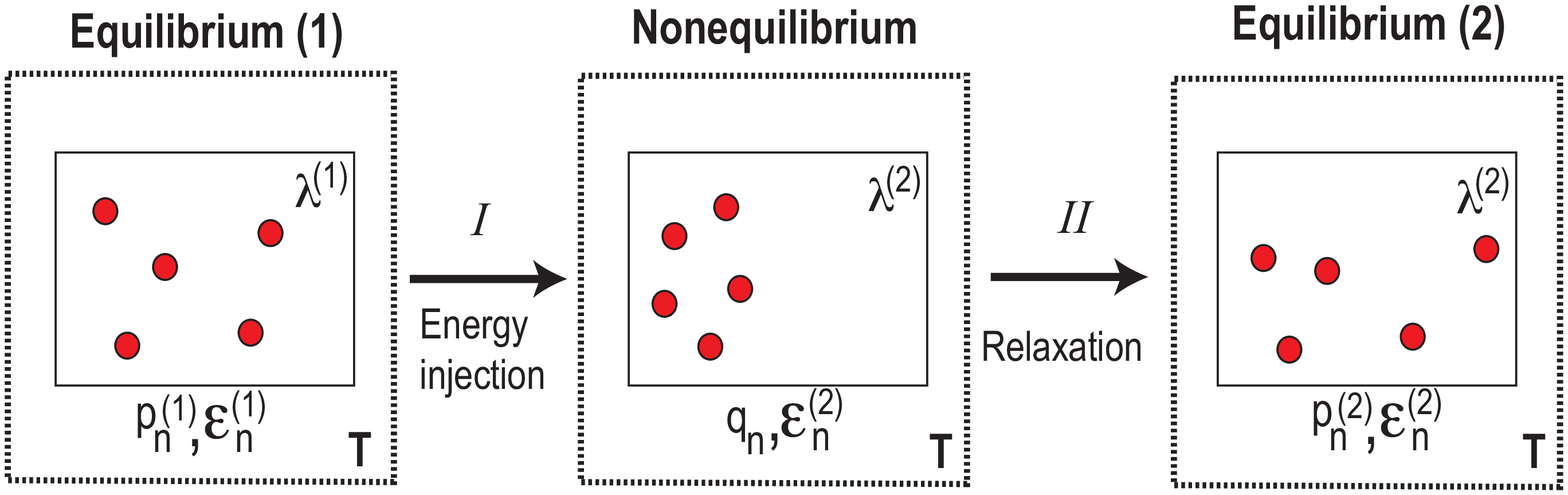}
\vspace{-2cm}
\caption{Generalized nonequilibrium quenches.  A system parameterized by $\lambda$ is coupled to an external bath at temperature $T$.  Initially, $\lambda= \lambda^{(1)}$ and the system is in equilibrium and is described by the Boltzmann distribution $p_n^{(1)}$.  $I$,  energy is suddenly injected into the system  while changing $\lambda$ from $\lambda^{(1)}$ to $\lambda^{(2)}$. The system is now described by  possibly nonthermal distribution, $q_n$.  During stage $II$ of the process, the system relaxes and equilibrates with the external bath after which it is described by a Boltzmann distribution $p_n^{(2)}$ with $\lambda=\lambda^{(2)}$.}
\end{figure}
During stage $I$, the total change in energy in the system $\Delta E^I$ can be divided into two parts,  adiabatic work, $W_{ad}^{I}$, and heat, $Q^{I}$,
\be
\Delta E^I= W_{ad}^I + Q^I.
\label{conEI}
\ee
 Adiabatic work is defined as the change in energy that would result from adiabatically changing the parameters from $\lambda^{(1)}$ to $\lambda^{(2)}$. Physically, it measures changes in total energy stemming  form  the parameter dependence of the  energy spectrum (potential energy). By definition, the heat is the remaining contribution to the change in energy \cite{fermi1936thermodynamics}. Thus in our language heat includes both the non-adiabatic part of the work and the conventional thermodynamic heat. The heat generated during process $I$ can be explicitly calculated (see Sec.~\ref{sec:derivation}):
\be
Q^I=T \Delta S^{I}+TS_r(q||p^{(2)})-TS_r(p^{(1)}||p^{(2)}),\quad \Delta S^{I}=S(q)-S(p^{(1)})
\label{def_QI}
\ee
where the $S(p) \equiv -\sum_n p_n \log p_n$ is the diagonal entropy of a probability distribution $p$ \cite{polkovnikov2011microscopic} and  
\[
S_r(q||p) \equiv \sum_n q_n \log(q_n/p_n)
\] 
is the relative entropy between the distributions $q$ and $p$. We have shown previously that for large ergodic systems, the diagonal entropy is equivalent to the usual thermodynamic entropy \cite{polkovnikov2011microscopic}. Note that for a cyclic process where $\lambda^{(1)}=\lambda^{(2)}$ the last term in Eq.~(\ref{def_QI}) vanishes since $p^{(1)}=p^{(2)}$. During process $II$,  the system re-equilibrates with the bath by exchanging heat, $Q^{II}$, with the reservoir. One can show that (see Sec.~\ref{sec:derivation}) 
\be
Q^{II} = T\Delta S^{II} - TS_r(q||p^{(2)}), \quad \Delta S^{II}=S(p^{(2)})-S(q).
\label{def_QII}
\ee
The importance of relative entropy for describing relaxation of nonequilibrium distributions has been discussed in previous for different setups both in quantum and classical systems \cite{levine1978information,schlogl1980, deffnerlutz2011,qian2001relative,mehta2008nonequilibrium}.  Taken together, (\ref{conEI}), (\ref{def_QI}), and (\ref{def_QII}) constitute the nonequilibrium identities that will be exploited next to calculate bounds for the efficiency of  engines that operate with a single heat bath.

\section{Maximum Efficiency of Engines}  
\label{sec:effic}

Figure 2 summarizes the single-reservoir engines analyzed in this work and compares them with Carnot engines~(\ref{carnot_engine}). The engine is initially in equilibrium with the environment (bath) at a temperature $T_0$ and the system is described by the equilibrium probability distribution $p_{eq}$. In the first stage, excess energy, $\Delta Q$, is suddenly deposited into the system. This can be a pulse electromagnetic wave, burst of gasoline, current discharge etc.  In second stage, the engine converts the excess energy into work and reaches mechanical equilibrium with the bath . Finally, the system relaxes back to the initial equilibrium state. Of course splitting the cycle into three stages is rather schematic but it is convenient for the analysis of the work of the engine. Such an engine will only work if the relaxation time of the system and environment is slow compared to the time required to perform the work. Otherwise the energy will be simply dissipated to the environment and no work will be done (see discussion in Ref.~[\onlinecite{LL5}]).

\begin{figure}
\includegraphics[width=15cm]{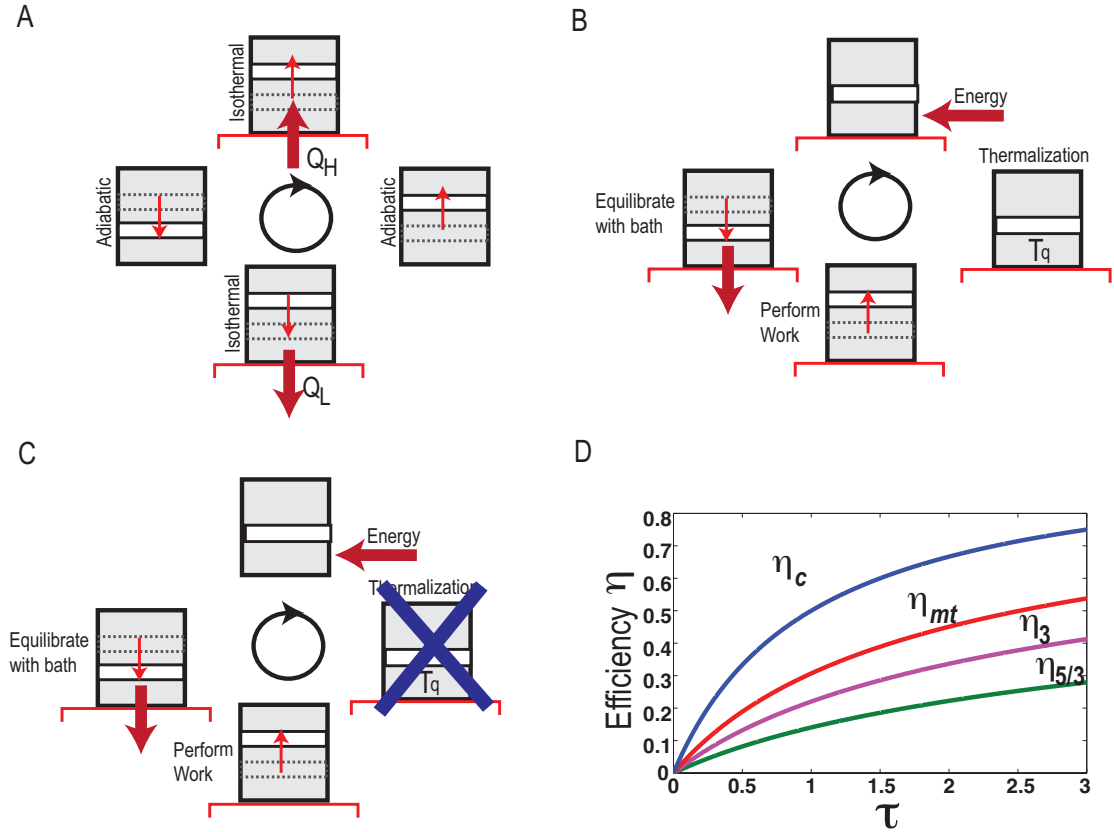}
\label{noneq_eng}
\caption{Comparison of Carnot engines and single-heat bath engines {\bf(A)} Carnot engines function by using two heat reservoirs, a  hot reservoir that serves as a  source of energy and a cold reservoir that serves as an entropy sink. {\bf(B)} In the ergodic regime, energy is injected into the engine. The gas within the engine quickly equilibrates with itself. The gas then performs mechanical work and then relaxes to back to its initial state. {\bf(C)}. In the non-ergodic regime, the system thermalizes on time scales much slower than time scales on which work is performed. {\bf(D)}. (blue) Maximum efficiency as a function of excess energy (ratio of injected energy to initial energy), $\tau$, for Carnot engine, $\eta_{c}$, (red) true thermodynamic bound, $\eta_{mt}$,  (magenta) actual efficiency of a non-ergodic engine which acts as an effective one-dimensional gas, $\eta_3$ (see the text), and (green) actual efficiency of three-dimensional ideal gas Lenoir engine, $\eta_{5/3}$.}
\end{figure}

 The initial injection of energy, $\Delta Q$ results in the corresponding entropy increase $\Delta S^{I}=S(q)-S(p_{eq})$ of the system, where $S$ is the diagonal entropy and $q$ describes the system immediately after the addition of energy. Because by assumption the environment is not affected during this initial stage, the total entropy change of the system and environment is also just $\Delta S^{I}$. By the end of the cycle, the entropy of the system returns to its  initial value. Thus, from the second law of thermodynamics, the increase in entropy of the environment must be greater than equal to $\Delta S^{I}$. This implies that the minimal amount of heat that must be dissipated into the environment during the cycle is $T_0\Delta S^{I}$. An engine will work optimally if no extra entropy beyond $\Delta S^{I}$ is produced during the system-bath relaxation since then all of the remaining energy injected into the system is converted to work. Thus, the maximal work that can be performed by the engine during a cycle is $W_{m}= \Delta Q- T_0 \Delta S^{I}$. For a cyclic process such as the one considered here, substituting (\ref{def_QI}) into the expression for $W_{m}$ implies that the maximum efficiency of a nonequilibrium engine, $\eta_{mne}$, is given by
 \be
 \eta_{mne}= \frac{W_m}{\Delta Q} = 1-\frac{T_0 \Delta S^{I}}{\Delta Q}= \frac{T_0S_r(q||p)}{\Delta Q}.
\label{efficiency_noneq}
 \ee 
Equation (\ref{efficiency_noneq}) is the main result of this paper. It relates the maximum efficiency of an engine to the relative entropy of the intermediate nonequilibrium distribution and the equilibrium distribution. We next consider various limits and applications of this result. We point that Eq.~(\ref{def_QI}) also allows us to extend the maximum efficiency bound to a more general class of engines, like Otto engines, where during the first stage of the cycle one simultaneously changes the external parameter $\lambda$ from $\lambda^{(1)}$ to $\lambda^{(2)}$. In this case,
 \be
 \eta_{mne}= \frac{T_0\left[S_r(q||p^{(2)})-S_r(p^{(1)}||p^{(2)})\right]}{\Delta Q},
\label{efficiency_noneq1}
 \ee 
where $p^{(1)}$ and $p^{(2)}$ stand for equilibrium Gibbs distributions corresponding to the parameters $\lambda^{(1)}$ and $\lambda^{(2)}$ at the beginning and the end of the process I respectively. Since the second term is negative, changing the external parameter during the first stage can only reduce the engine efficiency, though this may be desirable for other practical reasons unrelated to thermodynamics.

\subsection{Efficiency of Ergodic Engines}
An important special case of our bound is the limit where the the relaxation of particles within the engine is fast compared to the time scale on which the engine preforms work (see Figure 2). This is the normal situation in mechanical engines based on compressing gases and liquids. In this case, after the injection of energy the particles in the engine quickly thermalize and can be described by a gas at an effective temperature $T(E) \equiv (dS/dE)^{-1}$ that depends on the energy of the gas. It is shown in Sec.~\ref{sec:derivation_effic}, that in this case, (\ref{efficiency_noneq}) reduces to
\be
\eta_{mt}=1-{T_0\Delta S^{I}\over \Delta Q}={1\over \Delta Q}\int_E^{E+\Delta Q} dE'\left(1-{T_0\over T(E')}\right).
\label{eff_therm}
\ee
By definition $\eta_{mt}$ is the true upper bound for thermal efficiency of a single reservoir engine. 

It is easy to see that $\eta_{mt}$ is the integrated Carnot efficiency and thus it is always smaller that the Carnot efficiency corresponding to the same heating $Q_h=\Delta Q$ (see Fig. 2). This efficiency bound becomes very simple for ideal gases where $T(E)\propto E$. Assuming that in the beginning of the cycle the system is in equilibrium with environment, one has that the maximal efficiency of an equilibrium engines that thermalizes is 
\be
\eta_{mt}=1-{1\over \tau}\log(1+\tau),
\label{eff_id_gas}
\ee
where $\tau=\Delta Q/E=\Delta T/T_0$. For comparison the equivalent Carnot efficiency is 
\be
\eta_c={\tau\over \tau+1}.
\ee
It is interesting that the result for $\eta_{mt}$ is valid for arbitrary ideal gases and does not depend on dimensionality or the type of dispersion (linear, quadratic etc.) or the number of internal degrees of freedom. It is also valid for mixtures of ideal gases with different masses and dispersion relations. The expression~(\ref{eff_id_gas}) can be extended to the situations where the initial temperature of the engine is different from that of the environment (see Sec.~\ref{sec:derivation_effic}).

\subsection{Higher efficiency bound for non-ergodic distributions.} 

Another interesting limit is when the full thermalization time in the system is long compared to the time required to perform the work. We call engines that work in these parameter regime non-ergodic engines. This situation can be realized in small systems, integrable or nearly integrable systems with additional conservation laws or the systems where different degrees of freedom are weakly coupled like e.g. kinetic and spin degrees of freedom of molecules, electrons and phonons in metals and semiconductors and so on. In such systems the process of relaxation typically occurs in two stages. The system first undergoes a fast relaxation to a  quasi steady-state, prethermalized distribution. Subsequently, the system then very slowly relaxes to the true equilibrium distribution. The notion of prethermalization mechanism was first suggested in the context of cosmology~\cite{berges2004prethermalization}. Since it has been confirmed to occur both experimentally and theoretically in many physical situations including one and two dimensional turbulence~\cite{gurarie1995probability}, weakly interacting fermions~\cite{moeckel2010crossover}, quenches in low dimensional dimensional superfluids~\cite{gring_12} (see Ref.~[\onlinecite{polkovnikov2011colloquium}] for additional examples). 

Prethermalization is well known from standard thermodynamics where two or more weakly coupled systems first quickly relax to local equilibrium states and then slowly relax with each other. From a microscopic point of view, prethermalization is equivalent to dephasing with respect to a fast Hamiltonian $H_0$ where the density matrix effectively becomes diagonal with respect to the eigenstates of $H_0$. It was also recently realized that thermalization can be also understood as dephasing with respect to the full Hamiltonian of the system through the eigenstate thermalization hypothesis~\cite{deutsch1991quantum,srednicki1994chaos,rigol2008thermalization}. In the language of kinetic theory of weakly interacting particles, prethermalization implies  a fast loss of coherence between particles  governed by the noninteracting Hamiltonian $H_0=\sum_j m v_j^2/2$ followed by a much slower relaxation of the non-equilibrium distribution function to the Boltzmann form due to small interactions. 

\begin{figure}[ht]
\includegraphics[width=15cm]{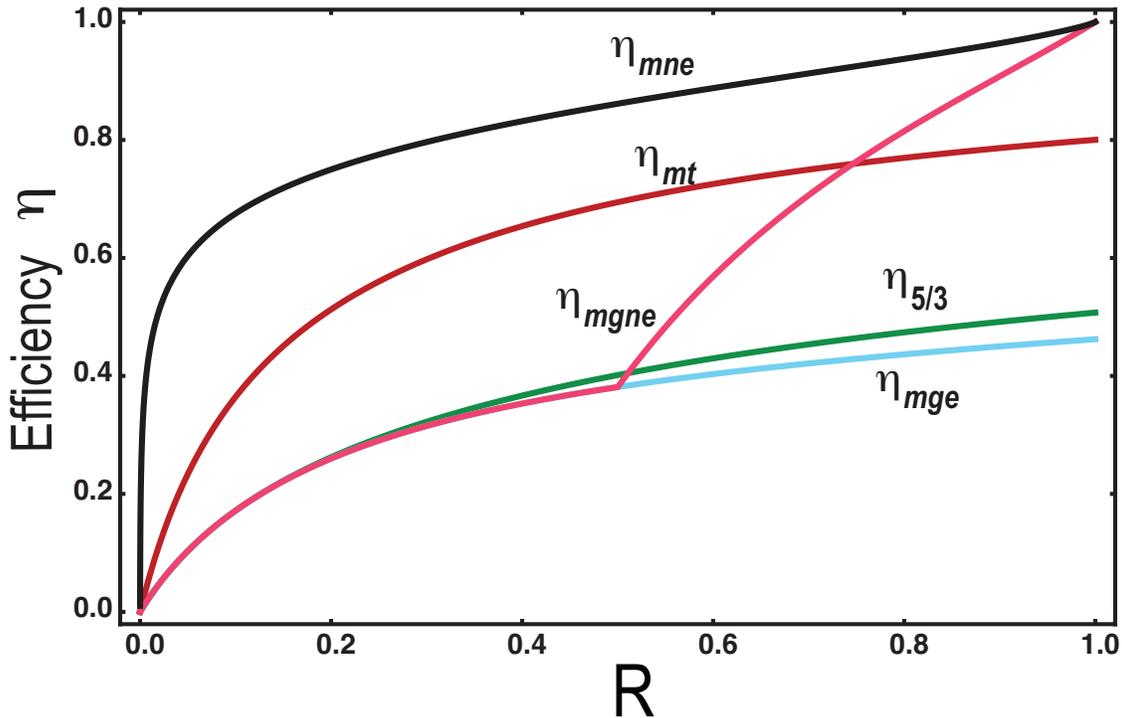}
\label{fig:eta_mag}
\caption{Efficiency of magnetic engine with initial temperature equal to $10\%$ of the Zeeman energy: $T_0=0.1 h_z$ as a function of the spin flipping rate (see Sec.~\ref{sec:examples}). (black)  $\eta_{mne}$, maximum non-ergodic efficiency, (red)  $\eta_{mt}$, maximum ergodic efficiency, (pink) $\eta_{mgne}$, efficiency of magnetic gas engine in non-ergodic regime,  (green)  $\eta_{5/3}$, efficiency of ergodic ideal gas engine,(blue) $\eta_{mge}$, efficiency of ergodic magnetic gas engine.}
\end{figure}

The efficiency of a non-ergodic engine is given by (\ref{efficiency_noneq}) with $q$ now representing the prethermalized distribution. A simple minimization shows that the numerator of (\ref{efficiency_noneq}) for a fixed energy increase $\Delta Q$ has a minimum precisely for the Gibbs distribution (see Sec.~\ref{sec:derivation_effic}). Thus, any non-equilibrium state can only increase the maximum possible efficiency of the engine.  Alternatively this statement can be understood from the fact that the Gibbs distribution maximizes the entropy for a given energy~\cite{jaynes1957information}. Thus for thermalizing engines the unavoidable amount of heating of the environment is maximum. Finally, notice that the first equality in  (\ref{efficiency_noneq}) implies  that the maximum value of $\eta_{mne}$, which is unity,  is achieved for a process   when the prethermalized non-equilibrium state has the same diagonal entropy as the initial state i.e where the probabilities $q_n$ are permutations of the probabilities $p_n$.  Thus, in principle, it is possible to create a non-ergodic heat engine, with efficiency arbitrary close to unity even if it is incoherent. We discuss an example of such an engine in the next section (see Fig.~\ref{fig:eta_mag} and related discussion).

\subsection{Maximum efficiency of coherent non-ergodic engines.}  Finally we briefly discuss the efficiency bound for coherent engines which preserve  coherence between particles while performing macroscopic work. Such engines are sensitive not only to conserved or approximately conserved quantities (like energy and velocity distribution for weakly interacting gas) but also to non-conserved degrees of freedom (like precise positions of particles at a given moment of time). In practice, such engines can be realized only for very small, non-interacting systems with long coherence times or in the systems where some macroscopic degrees of freedom are decoupled from the rest, like e.g. center of mass motion in solids. Such engines have the highest efficiency bound still given by Eq.~(\ref{efficiency_noneq}) but with $S_r$ standing for the full  relative entropy of the non-equilibrium distribution $q$ with respect to the equilibrium distribution $p$ (see Sec.~\ref{sec:derivation}):
\be
S_r^{vn}(q||p)={\rm Tr} \left[\rho\,(\log(\rho)-\log(p))\right].
\ee 
We will not further discuss such engines since they are not thermal. We only point that this bound based on relative von Neumann's entropy explains why mechanical engines can have arbitrary high efficiency. Indeed the von Neumann's entropy of a system of particles does not change if they start moving collectively implying that the bound given by~Eq.~(\ref{efficiency_noneq}) can reach unity.

\section{Some simple examples} 
\label{sec:examples}

\subsection{Ideal Gas Engine}

\subsubsection{Ergodic engine}

Let us start from the simplest ideal gas  single reservoir engine which pushes the piston. The engine undergoes the Lenoir cycle as illustrated in Figure 2.  First a pulse of energy $\Delta Q$ is deposited to the gas via e.g. a gasoline burst. The gas immediately thermalizes at a new temperature corresponding to the energy $E+\Delta Q$ and a new pressure. Then the gas undergoes adiabatic expansion pushing the piston and performing work until the pressure drops to the atmospheric value and finally the system relaxes back to the initial state at a constant pressure as the atmosphere pushes the piston back. Practically the engines based on the Lenoir cycle are not very efficient due to reasons unrelated to thermodynamics. We will use this cycle for illustration of our results because it is conceptually the simplest single heat reservoir engine.

The Lenoir cycle consists of three processes. Initially, the gas has pressure $P_0$, volume, $V_0$, and temperature $T_0$. Next, energy is injected at constant  volume so the effective temperature and pressure must rise. So after energy deposition, the system is described by pressure, $P_H$, volume $V_0$, and temperature $T_H$. The system then performs work by adiabatically expanding until the pressure equalizes. The system is then described by pressure, $P_0$, a volume $V_*$, and a temperature $T_*$. Finally the system relaxes back to the initial state by dropping temperature and volume back to $T_0$ and $V_0$ and a constant pressure $P_0$. To calculate the efficiency, we calculate the work the system performs and divide by the total heat added:
\be
\eta= \frac{W}{\Delta Q}.
\ee 
Denote the heat capacity ratio of an ideal gas by $\gamma=C_p/C_v$. It is related to the number of degrees of freedom $f$ per molecule by $\gamma=(f+2)/f$. We can write
\be
\Delta Q= \frac{f}{2} nR (T_H-T_0)= \frac{1}{\gamma-1} nR(T_H-T_0).
\label{defdeltaQIG}
\ee
By definition, the work is
\be
W=  \int_{V_0}^{V_*} (P(V)-P_0) dV,
\ee
To calculate the work during the adiabatic expansion, we use the fact that for an adiabatic process  the product $PV^{\gamma}$ is a constant. Thus, we can rewrite the equation above as
\begin{align}
W &=  P_0 V_*^\gamma \int_{V_0}^{V^\ast} dV\frac{1}{V^{\gamma} } - P_0(V_* -V_0) \\
\end{align}
Explicitly performing the integral yields,
\be
W =  \frac{P_0 V_*}{ \gamma-1}\left(\left(\frac{V_*}{V_0}\right)^{\gamma-1}-1 \right)- P_0(V_*-V_0).
\ee
We now use the relation $P_*V_*^{\gamma}=P_HV_0^{\gamma}$ and the ideal gas law to find $V_*=V_0(1+\tau)^{1/\gamma}$, where $\tau=(T_H-T_0)/T_0$. Then
\be
W=  P_0 V_0 \left[{\tau\over \gamma-1}-{\gamma\over \gamma-1}\left((1+\tau)^{1/\gamma}-1\right)\right].
\ee
Finally rewriting Eq.~(\ref{defdeltaQIG}) as $\Delta Q={\tau\over \gamma-1}P_0 V_0$ we find:
\be
\eta_\gamma=1-{\gamma\over \tau}\left( (1+\tau)^{1/\gamma}-1\right).
\label{eta_gamma}
\ee
 The efficiency $\eta_\gamma$ is bounded by the maximum thermodynamic efficiency~(\ref{eff_id_gas}) as it should approaching this bound as $\gamma\to\infty$ and for $\gamma\to 1$ (minimal possible value) the maximum efficiency $\eta_\gamma$ goes to zero. For a monoatomic gas we have $\gamma=5/3$ and the corresponding efficiency is plotted in Fig. 2.  For a typical value $\tau=1$ where the temperature increases by a factor of 2 during the pulse for monoatomic gas we find $\eta_{mt}\approx 0.31$ and $\eta_\gamma\approx 0.14$, i.e. the efficiency of such engine is significantly below the thermodynamic bound (which in turn is considerably less than the Carnot bound $\eta_c=0.5$). For $\tau=2$, i.e. when the temperature jumps by a factor of three the situation is somewhat better $\eta_{mt}\approx 0.45$ while $\eta_\gamma\approx 0.22$. For more complicated molecules with $\gamma$ closer to one the efficiency is even less.

\subsubsection{Non-ergodic Engine}

We now analyze performance of a non-ergodic ideal gas engine of the following form. Consider, the scenario where an energy pulse generates a fraction of very fast particles moving horizontally, which very slowly thermalize with the rest of the particles. In this case these particles can be treated as effectively a one dimensional gas with $\gamma=3$ such that Eq.~(\ref{eta_gamma}) applies.   Microscopically this result can be understood by using the conservation of the adiabatic invariants~\cite{LL1}. Indeed, during the slow motion of the piston the fast particles approximately conserve adiabatic invariants equal to the product of the momentum, $p$, and twice the distance between piston and the wall, which we denote $V$ (since in our setup the area of the piston does not change the length and the volume aree equivalent). This implies
 \be
 pV= C_1,
 \ee
 with $C_1$ a constant. Thus, we expect that
 \be
 p \propto V^{-1}
\ee
Furthermore, consider the pressure, $P$, of such a gas can be thought of as the force per unit area or equivalently the energy density per unit volume,
\be
P \propto p^2 /V
\ee
Taken, together these relations imply that
\be
PV^3={\rm const}.
\ee
 This is precisely the relationship for a gas adiabatically expanding with $\gamma=3$. Thus, the efficiency is equivalent to that of an adiabatic 1D gas with $\gamma=3$. The efficiency of this non-ergodic engine is still below thermodynamic bound $\eta_{mt}$ because the latter does not depend on dimensionality, but it is much higher than the efficiency of the ergodic engine according to our general expectations (see Fig. 2). In particular $\gamma_3(1)\approx 0.22$ and $\gamma_3(2)\approx 0.33$ i.e. we are getting approximately 50$\%$ improvement of the efficiency compared to the ergodic gas. With this simple design it is impossible to exceed the thermodynamic bound $\eta_{mt}$ because pressure is only sensitive to the overall kinetic energy not to details of the energy distribution.

Interestingly $\eta_\gamma$ given by Eq.~(\ref{eta_gamma}) also describes the efficiency of a photon engine where the piston is pushed by the photon pressure created by some light source like a bulb. In this case one should use $\gamma=4/3$ in the ergodic case, where the photon gas is equivalent to the black-body radiation at a higher temperature and $\gamma=2$ in the non-ergodic case where the photons are effectively one-dimensional. Again the non-ergodic setup allows one to increase the engine efficiency.

\subsection{Magnetic Gas Engine}

It is possible exceed the thermodynamic efficiency $\eta_{mt}$ by considering more complicated engines with an additional magnetic degree of freedom.   Then as we show below one can create a non-ergodic engine with efficiency higher than the thermodynamic bound and which can be arbitrarily close to $100\%$. Assume that we have a gas composed of $N$ atoms which have an additional magnetic degree of freedom like a spin. For simplicity we assume that the spin is equal to $1/2$, i.e. there are two magnetic states per each atom. As  will be clear from the discussion,  this assumption is not needed for the main conclusion and the calculations can easily be generalized to the case where we consider electric dipole moments or some other discrete or continuous internal degree of freedom instead of the spin.

The Hamiltonian of the system is then
\be
\mathcal H_0=\sum_j {m v^2_j\over 2}-h_z\sigma^z_j,
\ee
where $\sigma^z_j$ are the Pauli matrices.  To simplify notations we absorbed the Bohr magneton and the $g$ factor into the magnetic field. The first term in the Hamiltonian is just the usual kinetic energy and the second term is due to the interaction of the spin degrees of freedom with an external field in the $z$-direction.  Initially the system is in equilibrium at a temperature $T$ and a fixed magnetic field $h_z$.

Now let us assume that via some external pulse we pump energy to the atoms by flipping their spins with some probability. This can be done by a resonant laser pulse or by e.g. a Landau-Zener process where we adiabatically turn on a large magnetic field in $x$-direction then suddenly switch its sign and slowly decrease it back to zero. Ideally this process creates a perfectly inverse population of atoms (i.e. number of spin up and spin down particles is exchanged) but in practice there will be always some imperfections. In general unitary process the new occupation numbers can be obtained from a single parameter describing the flipping rate: $R\in [0,1]$, with
\be
q_\uparrow=p_\uparrow (1-R)+p_\downarrow R,\; q_\downarrow=p_\downarrow (1-R)+p_\uparrow R
\label{defqmg}
\ee
During such a process, the energy added to the system is 
\be
\Delta Q=2N \mu h_z R (p_\uparrow-p_\downarrow)=2N \mu h_z R \tanh {\mu h_z\over T}.
\label{defQmg}
\ee
As expected this energy is non-negative.  As before we will discuss first the ergodic and then the non-ergodic engines.

\subsubsection{Ergodic Engine}
In the equilibrium ergodic case the atoms are first allowed to relax to a thermal distribution corresponding to the new energy. This will result in a higher effective temperature $T_H$ for the magnetic gas. This temperature can be found from the equation relating temperature to energy:
\be
{3T_H\over 2}-N\mu h_z\tanh{\mu h_z\over T_H}=E+\Delta Q,
\ee
where $E$ is the initial equilibrium energy of the system and $\Delta Q$ is found in Eq.~(\ref{defQmg}).

Work can be extracted in a similar manner to the ideal gas engine considered above by letting the gas adiabatically expand and push a piston until the pressures equilibrate. We know that the work done during such a process is 
\be
W= \int_{V_0}^{V_*} (P(V)-P_0)dV,
\ee
where we have adapted the notation of the last section.

 During an adiabatic expansion the entropy must be conserved. The entropy as a function of the volume and temperature of magnetic gas is given by
 \be
{S(T, V)\over N}=C+\log(V)+{\log(T)\over \gamma-1}+\log\left[2\cosh{\mu h_z\over T}\right]-{\mu h_z\over T}\tanh{\mu h_z\over T},
\ee
where $C$ is an unimportant constant and $V$ is the volume. Notice that the entropy has contributions from both the kinetic and magnetic sectors. Additionally, we know that for the gas,
\be
P(V,T)= NT(V)/V,
\ee
where the temperature, $T(V)$, is now considered a function of volume during the adiabatic expansion. $T(V)$ can be solved for from the  self-consistency condition for adiabatic expansion
\be
S(T(V), V)= S(T_H, V_0).
\ee
Together, these relations allow us to numerically solve for the work performed by the engine during adiabatic expansion and the results are shown in Figure 3. Notice, that the additional spin contribution to the entropy makes the engine somewhat less efficient than the ideal gas engine because the additional entropy is eventually released in the form of heat. However, this difference can be very small if the initial temperature is small compared to the Zeeman energy splitting (see Figure 3).

\subsubsection{Non-ergodic Engine}

In the non-ergodic setup spins are allowed to do work before they relax with kinematic degrees of freedom. The easiest setup we can imagine is to rotate the spins around the $x$ axis by the angle $\pi$ when $R>1/2$, i.e. when there is an inverse spin population. This can be done by e.g. applying a strong magnetic field along the $x$-axis for exactly half the period of the Larmor precession. This extra field does not do any work by itself since the magnetization is orthogonal to the $x$-axis. We can extract  a magnetic work $W_{mag}$ from the system by coupling the magnetization of the spins to the source of the external z-component of the magnetic field. The amount of  ``magnetic''  work generated from such a device is
\be
W_{mag}= N |M|h_z=N(2 q_\downarrow-1)
\ee
Note that the same maximum work can be extracted from the system in the form of a coherent light pulse. The remaining excess energy can be used as in the previous part by allowing the gas to adiabatically expand and drive a piston. Thus, an additional work, $W_{ad}$ can be extracted by using the equilibrium equations from the last section with $\Delta Q \rightarrow \Delta Q- W_{mag}$  Note that for $R>1/2$ we always have $\Delta Q>W_{mag}$.  This allows us to calculate the efficiency for this system using the epxression
\be
\eta = \frac{W_{mag}+ W_{ad}}{\Delta Q}.
\ee
The resulting efficieny  is plotted in Figure 3. This efficiency, as it should, is always bounded by the maximum nonequilibrium efficiency given by the relative spin entropy:
\be
\eta_{mne}= \frac{TS_r(q||p)}{\Delta Q}={TN \left[q_\uparrow \log\left({q_\uparrow\over p_\uparrow}\right)+q_\downarrow \log\left({q_\downarrow\over p_\downarrow}\right)\right]\over \Delta Q}
\ee
with $q_{\uparrow,\downarrow}$ given by $(\ref{defqmg})$ and $\Delta Q$ given by $(\ref{defQmg})$. Note that as the flip rate $R$ approaches unity, i.e. the up and down spins exactly exchange, the efficiency of this engine approaches unity. This is related to the fact that such a pulse does not generate the additional entropy in the system and once the spins are rotated by half the Larmor period around $x$-axis performing the macroscopic work they are in equilibrium state with no additional relaxation required. It is easy to see that any imperfections like $R<1$ or small disorder in spin Larmor frequencies will decrease the engine efficiency. because there is always excess magnetization and excess entropy compared to the initial equilibrium state.

\section{Derivation of Thermodynamic Identities}
\label{sec:derivation}

 In this section, we derive the nonequilibrium fundamental thermodynamic relations established earlier As shown in Figure 1, we consider a process that proceeds in two parts: In part I), the system $A$ is prepared in equilibrium and is characterized by a Gibbs distribution with temperature $T$ and parameters $\lambda=\lambda^{(1)}$. We denote the initial distribution by $p_n^{(1)}$. The system $A$ then undergoes an arbitrary process, which brings it to the new possibly nonequilibrium state while simultaneously changing $\lambda$ from $\lambda^{(1)}$ to $\lambda^{(2)}$. Since the energy distribution only depends on the probabilities of occupying the eigenstates $q_n$, which form the so called diagonal ensemble, we will be concerned only by these probabilities. The distribution $q_n$ can correspond to an effective thermal distribution at a higher temperature, or  to a prethermalized, nonequilibrium distribution.  For example for weakly interacting gas of particles $q_n$ is described by their possibly nonequilibrium momentum distribution.  In stage II) the system $A$ prepared in the nonequilibrium state relaxes to a new equilibrium state with the bath $B$ and is described by an equilibrium distribution $p_n^{(2)}$ with $\lambda=\lambda^{(2)}$. In principle during this relaxation one can still perform the work on $A$ but for simplicity we assume this does not happen.

We start by defining several important concepts. The first is the diagonal entropy of a distribution, $p_n$, which we label $S(p)$. The diagonal entropy  is the same as the Shannon entropy of the distribution and reduces to the thermodynamic entropy for large systems  \cite{polkovnikov2011microscopic}. It is explicitly given by,
\be
S(p)= -\sum_n p_n \log{p_n}.
\ee 
 For stationary distributions the diagonal entropy is the same as the von Neumann's entropy $S_{vn}(\rho)=-{\rm Tr}\,[ \rho \log(\rho)]$, with $\rho$ the corresponding density matrix. A second related concept is the relative entropy, $S_r(q||p)$ between two distributions $p$ and $q$, which is defined as 
\be
S_r(q||p) =\sum_n q_n \log{\frac{q_n}{p_n}}.
\ee
 In general, the notion of relative entropy can be extended to full density matrices, $\rho$ and $\rho_q$, rather than diagonal parts: 
\be
S^{vn}_r(\rho_q||\rho)={\rm Tr}\,[ \rho_q (\log(\rho_q)-\log(\rho_p))]
\label{svn_rel}
\ee 
As we already mentioned in this work we will be interested in only the diagonal part of the distribution $q$ and thus in the associated entropy $S_r$ and not $S_r^{vn}$.

Another related concept is the adiabatic work, $W_{ad}$ (see also a related discussion in Ref.~[\onlinecite{polkovnikov2008heat}]). Consider a system parameterized by $\lambda$ with a $\lambda$-dependent energy spectrum, $\mathcal E_n^{\lambda}$. Lets assume that the system is initially in equilibrium with a bath at temperature $T$ described by a Boltzmann distribution
\be
p_n^{(1)}= \frac{e^{-\beta \mathcal E_n^{(1)}}}{Z^{(1)}}
\ee
with $Z^{(1)}=\sum_n e^{-\beta \mathcal E_n^{(1)}}$ the usual partition function. Now consider a process where an external parameter is adiabatically changed from $\lambda= \lambda^{(1)}$ to $\lambda=\lambda^{(2)}$. Since this is done adiabatically, the probability distribution $p_n^{(1)}$ does not change during the process.  We define the total change in energy of the system during such an adiabatic process the adiabatic work, $W_{ad}$. It represents the \emph{minimum} amount of work that can be done in changing parameter from $\lambda^{(1)}$ to $\lambda^{(2)}$ and is given by
\be
W_{ad}^I= \sum_n p_n^{(1)} (\mathcal E_n^{(2)}- \mathcal E_n^{(1)}).
\ee

After defining these concepts let us consider process $I$. The total energy change during this process is
\be
\Delta E^I = \sum_n q_n \mathcal E_n^{(2)} -p_n^{(1)} \mathcal E_n^{(1)}.
\ee
Now notice that we can rewrite
\begin{multline}
\Delta E^I  = W_{ad}^I + \sum_n (q_n -p_n^{(1)}) \mathcal E_n^{(2)} = W_{ad}^I-T \sum_n (q_n - p_n^{(1)}) \log{p_n^{(2)} }= W_{ad}^I\\
 +T\sum_n q_n \log{\frac{q_n}{p_n^{(2)}}} -T \sum_n p_n^{(1)} \log{\frac{p_n^{(1)}}{p_n^{(2)}}} +T \sum_n p_n^{(1)}\log{p_n^{(1)}} -T \sum_n q_n \log{q_n} = W_{ad}^I + S_r(q||p^{(2)})\\- S_r(p^{(1)}||p^{(2)}) + T[S(q)-TS(p^{(1)})]
= W_{ad}^I+ S_r(q||p^{(2)})- S_r(p^{(1)}||p^{(2)}) + T \Delta S^I
\end{multline}.
By definition, we have that the heat $Q^I$ is just
\be
Q^I =\Delta E^I -W_{ad}^I=  T \Delta S^I+TS_r(q||p^{(2)})-T S_r(p^{(1)}||p^{(2)}) 
\label{first}
\ee
This is the first of the thermodynamic entities. For a cyclic process or a standard heating process where the parameter $\lambda^{(2)}=\lambda^{(1)}$ we have $p^{(1)}=p^{(2)}$ and the last term vanishes so
\be
Q^I =\Delta E^I -W_{ad}^I =  T \Delta S^I+TS_r(q||p^{(2)})
\ee
Because the relative entropy is non-negative we immediately see that $Q^{I}-T\Delta S^{I}\geq 0$, which is the first inequality in Eq. (3). We emphasize again that here $T$ is the initial temperature of the system.

In process $II$ when the system relaxes, the total change in energy is by definition the heat, $Q^{II}$, exchanged with the reservoir. In this case, we can write
\begin{multline}
Q^{II} =\sum_n (p_n^{(2)}-q_n) \mathcal E_n^{(2)}=  T\sum_n (q_n-p_n^{(2)}) \log p_n^{(2)}  + T \sum_n q_n \log q_n - T \sum_n q_n \log q_n \\
= T S(p^{(2)})-TS(q)-T S_r(q||p^{(2)})= T \Delta S^{II} -TS_r(q||p^{(2)})
\label{DeriveII}
\end{multline}
This yields the second thermodynamic identity, from which the second inequality in Eq. (3) immediately follows (now $T$ is the temperature of the bath where the initial distribution $q$ relaxes to). This result implies that the relative entropy between the arbitrary nonequilibrium and equilibrium distributions has a physical meaning of the total entropy generation in the system + bath during the relaxation of the nonequilibrium distribution to equilibrium. Indeed by energy conservation the heat dissipated to the bath is $Q_B=-Q^{II}$. Because the temperature of the bath does not change we can use the standard thermodynamic identity $Q_B=T \Delta S_B$. Combining this with Eq.~(\ref{DeriveII}) and changing the notation $\Delta S^{II}\to \Delta S_A$ we indeed find that
\be
S_r(q||p^{(2)})=\Delta S_A+\Delta S_B
\ee
As expected in accord with the second law of thermodynamics the total entropy change in the system and the bath is always non-negative as $S_r(q||p^{(2)})\geq 0$ for arbitrary $q$.

 Let us finally point that the relations (\ref{first}) and (\ref{DeriveII}) are also valid if we use the von Neumann's entropy  (in contrast to the Diagonal entropy) to define $\Delta S^{I,II}$ and the corresponding von Neumann's relative entropies given by Eq.~(\ref{svn_rel}). In this form the similar relations were obtained earlier in Ref.~[\onlinecite{deffnerlutz2011}], see also Ref.~[\onlinecite{schlogl1980}] for similar results in classical systems.

\subsection*{Quantum - classical correspondence}
\label{sec:entities}

In this section, we discuss the relationship between the``quantum'' notation used in the paper and the classical thermodynamic quantities that usually depend only on energies.  For macroscopic systems, we can replace a probability to be in a state $p_n$ by the probability, $W_p(E)$, that the system has energy $E$ by multiplying by an appropriate density of states, $\Omega(E)$ (see Ref.~[\onlinecite{bunin2011universal}]). We make the usual identifications 
\be
W_p(E) = p(E) \Omega(E)
\ee
and replace sums by integrals
\be
\sum_n  \rightarrow \int dE \Omega(E).
\ee

This identification allows us to translate all expression in the text to usual thermodynamic expressions. As an illustration consider the diagonal entropy,
\be
S= -\sum_n p_n \log p_n. 
\ee
Under the identification above this becomes
\be
S= -\int dE W_p(E) \log W_p(E) + \int dE  W_p(E) \log \Omega(E).
\ee
For macroscopic systems, we know that the density of states are well approximated by a Gaussian (see Ref.~[\onlinecite{bunin2011universal}])
\be
W_p(E) =\frac{ e^{-(E-\bar{E})^2/2\delta E^2}}{2 \pi \delta E^2}.
\ee
Therefore, we can calculate the entropy and one finds
\begin{align}
S(\bar{E}) &= \log{\sqrt{2\pi e} \delta E}+ \int dE  W_p(E) \log \Omega(E) \\
&\approx \log(\sqrt{2\pi e} \,\delta E \Omega(\bar{E}))
\end{align}
As discussed in section II.D of  Ref.~[\onlinecite{polkovnikov2011microscopic}] this is precisely the thermodynamic entropy. We note that there is a recent different derivation of this result based on the saddle point approximation~\cite{gurarie2012entropy}.

\section{Maximum Efficiency of Thermal Engines}
\label{sec:derivation_effic}

\subsection{Ergodic engines: general thermodynamic considerations}
We will start our analysis from the most straightforward setup where the system is macroscopic and the relaxation time within the system is the fastest time scale in the problem. This is the natural situation in gases, liquids and solids where the relaxation times are extremely fast. In this situation the initial energy increase $\Delta Q^I$ results in the corresponding entropy increase $\Delta S^I=S(E+\Delta Q)-S(E)$.  Because { by assumption the bath is not affected during this initial stage, this entropy increase equals to the total entropy change of the system and environment. For remainder of  the engine's cycle, the total entropy change of the system and environment must be non-negative due to the second law of thermodynamics. By the end of the cycle, the entropy of the system returns to its  initial value. Consequently, the entropy of the environment must increase by an amount $\Delta S^I$ or larger. Thus, the minimal heat dissipated to the environment during the cycle is $T_0\Delta S^I$ where $T_0$ is the temperature of the environment. Thus the maximum amount of work that can be performed during the cycle is $W_{max}=\Delta Q^I- T_0\Delta S^I$. Hence, the maximum thermodynamic efficiency of our engine is
\be
\eta_{mt}=1-{T_0\Delta S^I\over \Delta Q^I}={1\over \Delta Q^I}\int_E^{E+\Delta Q^I} dE'\left(1-{T_0\over T(E')}\right).
\label{eff_therm1}
\ee
where $T(E)=dE/dS$ is the equilibrium temperature of the system. We emphasize that deriving this result we never assumed that the initial energy change in the system is quasi-static. We have only used the fact that the equilibrium entropy is a unique function of energy. The result (\ref{eff_therm}) is very general since it is obtained with the single assumption of fast equilibration within the system. This assumption is justified for most practical heat engines like combustion engines. As discussed earlier, this efficiency bound becomes very simple for ideal gases where $T(E)\propto E$. Assuming that in the beginning of the cycle the system is in equilibrium with environment: $T(E)=T_0$, we recover the bound given by Eq.~(\ref{eff_id_gas}).

%\subsection{Initial temperature of engine not equal to environment}

The result (\ref{eff_therm}) still applies if in the beginning of the cycle the system is at some temperature $T_i\equiv T(E)$, which is not the same as the temperature of the environment.  This setup is e.g. realized in Otto engines, which are closest prototypes of real combustion engines. From the conceptual point of view this can happen if the engine does not have time to fully relax to equilibrium with the environment during one cycle. It is easy to see that in the limit when $(T(E+\Delta Q^I)-T_i\ll T_0$, i.e. when the excess impulse $\Delta Q^I$ does not substantially change the temperature of the engine, the maximum efficiency is given by the Carnot limit. Practically, this situation is probably  not advantageous since an engine which is hot most of the time will constantly radiate heat to the atmosphere. 

For ideal gases with $T_i\neq T_0$ (\ref{eff_therm}) generalizes to
\be
\eta_{mt}=1-{\tau_i\over \tau}\log(1+\tau),
\label{eff_id_gas1},
\ee
where $\tau_i=T_0/T_i$. When $\tau_i=1$ Eq.~(\ref{eff_id_gas1}) obviously reduces to Eq.~(\ref{eff_id_gas}).

\subsection{Explicit derivation from general expression}

In this section, we re-derive these results starting with the general expression
\be
\eta = \frac{TS_r(q||p_{eq})}{\Delta Q}.
\label{generalefficiency}
\ee
Rather than repeating and expanding the ``quantum'' derivation given earlier we will work here directly with continuous distributions to give the reader a feel of how these results can be derived directly from classical statistical physics.  In this derivation we will make no assumptions about the form distribution $q$ and assume that the equilibrium distribution $p(E)$ obeys the Bolztmann's form:
\be
p(E)={1\over Z}\exp[-\beta E],
\ee
where $Z$ is the continuous partition function.

The key assumption in our derivation is that after the initial pulse of energy the engine reaches a stationary state with respect to the fast Hamiltonian $H_0$, i.e. the relaxation with respect to $H_0$ is the fastest time scale in the problem. If $H_0$ is interacting and the system is large then such relaxation is equivalent to the thermalization in the engine at some higher temperature. But if $H_0$ is noninteracting or the engine is very small, consisting of very few degrees of freedom then such relaxation leads to some stationary nonequilibrium state. For integrable systems such state can be well described by a generalized Gibbs ensemble~\cite{rigol2007GGE}. For a single particle in a chaotic cavity such relaxation means loss of memory about the position and direction of the momentum for a particle moving along a periodic orbit in a regular cavity the relaxation implies loss of memory of the coordinate along the trajectory (see Ref.~[\onlinecite{bunin2011universal}] for additional discussion). The important for us mathematical result, which extends the Araki-Lieb's subadditivity theorem~\cite{arakilieb1970} to the diagonal entropy, is that if two systems are prepared in stationary states and then coupled in an arbitrary way the sum of their diagonal entropies can only increase or stay the same~\cite{polkovnikov2011microscopic}. Thus 
\be
\Delta S^{II}+\Delta S_B\geq 0.
\label{SIIB}
\ee
In the same work (see also Sec.~\ref{sec:entities}) it was also proven that for large systems like the bath the diagonal entropy reduces to usual thermodynamic entropy and that it obeys the fundamental thermodynamic relation: $T\Delta S_B=Q_B$. From the energy conservation we have $Q_B=-Q^{II}$ thus we find
\be
Q^{II}\leq T\Delta S^{II}.
\label{Q_inequality}
\ee
So the minimal amount of heat dissipated to the bath is equal to the difference between the diagonal entropies of the system in the nonequilibrium state $q$ and the equilibrium state. Therefore the maximum efficiency of any engine, equilibrium or nonequilibrium, is given by
\be
\eta_m={\rm max}\left[ {Q^I+Q^{II}\over Q^{I}}\right]={Q^I+T\Delta S^{II}\over Q^I}
\ee

Now let us evaluate these terms explicitly
\begin{multline}
Q^I=\int dE \Omega(E) E (q(E)-p(E))=\\
=\int dE\, E (W_q(E)-W_p(E))=T\int dE (W_p(E)-W_q(E))\log(p(E)).
\end{multline}
where $W_q(E)=\Omega(E)q(E)$ is the nonequilibrium energy distribution after the initial energy pulse. Similarly (see Ref.~[\onlinecite{polkovnikov2011microscopic}])
\be
T\Delta S^{II}=T\int dE [W_q(E)\log(q(E))-W_p(E)\log(p(E))]
\ee
Thus we find
\be
Q^{I}+T\Delta S^{II}=T\int dE\, W_q(E)\log\left[{q(E)\over p(E)}\right]=\int dE\, W_q(E)\log\left[{W_q(E)\over W_p(E)}\right]=S_r(q||p).
\label{QI}
\ee
This completes the proof of Eq.~(\ref{generalefficiency}).

%\subsection{A Brief Digression on Coherent Engines}
We can also generalize the result to ``coherent'' engines where the quantum coherence between particles is maintained.  As we explained in the main text such engines either work on time scales faster than relaxation even with respect to the fast Hamiltonian $H_0$ or which preserve coherence during evolution, e.g. where all particles perform a collective oscillatory motion. For this class of engines the proof above stands if we use the von Neumann's entropy instead of diagonal entropy for the system in Eq.~(\ref{SIIB}) and Eq.~(\ref{Q_inequality}) as well as the full relative entropy $S_r^{vn}(\rho_q||\rho)$ (see Eq.~(\ref{svn_rel})) instead of $S_r(q||p)$ in Eq.~(\ref{QI}). The proof relies on subbtivity of the von Neumann's entropy~\cite{arakilieb1970}, which for the system and the bath states $S_{AB}^{vn}(t)\leq S_A^{vn}(t)+S_B^{vn}(t)$. Using that $S_B^{vn}(t)\leq S_B(t)$~\cite{polkovnikov2011microscopic} and $S_{AB}^{vn}(0)=S_A^{vn}(0)+S_B(0)$ as well as the relation $T\Delta S_B=\Delta Q_B$ we prove Eq.~(\ref{QI}) with $S_r(q||p)\to S_r^{vn}(\rho_q|| \rho)$. It is easy to check that the relative entropy can only increase if the matrix $q$ becomes off-diagonal so as expected the engines which preserve coherence during work cycle can be more efficient than engines which loose coherence due to dephasing.

\subsection{Increased Efficiency of non-ergodic engines}

 The physical reason for inevitable losses in engines is the second law of thermodynamics which  states that the total entropy of the system and bath can only increase or stay the same. In the class of engines we consider, the entropy is generated during the initial pulse of energy and its dissipation to the bath leads to heating losses. Thus, it is intuitively clear that a engine can be made more efficient if the the entropy added to the system during the initial energy pulse is minimal. Because the equilibrium Gibbs distribution maximizes entropy for a given energy, it is clear that the nonequilibrium engines with smaller entropy have the potential to more efficient than equilibrium engines. Here we mathematically prove that this is indeed the case using the general result~(\ref{generalefficiency}). In particular, below we prove that for a given energy change $\Delta Q$ the relative entropy between $q$ and $p$ distributions is minimal when $q(E)$ is described by the Gibbs form. The minimum of the relative entropy can be found by extremizing the expression:
\be
\sum_n q_n [\log(q_n)-\log(p_n)]+\alpha_1 \sum_n E_n (q_n-p_n)+\alpha_2\sum_n q_n
\ee
with respect to the set of $q_n$, where $\alpha_1$ and $\alpha_2$ are the Lagrange multipliers enforcing the fixed energy and the probability conservation. Differentiating the result above with respect to a particular $q_m$ we find
\be
\log(q_m)+1-\log(p_m)+\alpha_1 E_m-\alpha_2=0,\; \Rightarrow\; q_m=C\exp[-(\beta+\alpha_1) E_m].
\ee
I.e. the distribution $q$, which  extremizes the relative entropy with respect to the Gibbs distribution is another Gibbs distribution corresponding to a different temperature fixed by the energy change. The fact that this extremum is minimum is obvious from considering the zero energy change case. This can be also checked explicitly by taking the second derivatives of the relative entropy with respect to the set of $q_n$.

\section{Discussion}

In this work, we extended the fundamental thermodynamic relation to a large class of nonequilibrium phenomena where energy is suddenly injected into a system coupled to a thermal bath. Using these relations, we have derived new thermodynamic bounds for engines that operate with a single reservoir. In particular, we have shown that the efficiency can be related to the relative entropy of the nonequilibrium distribution immediately following the injection of energy. Our new bound has several striking implications. First, we find that the true thermal efficiency of single-reservoir engines is below the corresponding Carnot bound. This gives a better metric for measuring the efficiency of actually existing engines. Second, the efficiency of engines can be increased by using nonequilibrium (prethermalized) distributions.  Taken together, this suggests some broad guidelines for building more efficient engines.  

The Carnot engine has served as a major source of intuition for increasing the efficiency of real engines. It is our hope that the nonequilibrium efficiency (\ref{efficiency_noneq}) can provide similar intuition for nonequilibrium engines. Of course, the technological challenge of constructing a specific engine which can harness  the additional efficiency possible with nonequilbrium engines remains an open question. Nonetheless, as illustrated with the simple examples considered here, this represents a tantalizing new possibility for future engine designs.

\begin{acknowledgments}
We would like to acknowledge Luca D'Alessio, Sebastian Deffner, Shyam Erramilli,  Yariv Kafri, and Clemens Neuenhahn for useful discussions and feedback on the paper. AP and PM were  partially funded by a Sloan Research Fellowship. Work of AP was funded by BSF 2010318, NSF DMR-0907039, AFOSR FA9550-10-1-0110, and the Simons Foundation.
\end{acknowledgments}


\begin{thebibliography}{10}

\bibitem{reif1965fundamentals}
Reif F
\newblock (1965) \emph{Fundamentals of Statistical and Thermal Physics
  (McGraw-Hill Series in Fundamentals of Physics)}
\newblock (McGraw-Hill Science/Engineering/Math).

\bibitem{fermi1936thermodynamics}
Fermi E
\newblock (1956, c1936) \emph{Thermodynamics}
\newblock (Dover).

\bibitem{magnasco1993forced}
Magnasco M
\newblock (1993) Forced thermal ratchets.
\newblock \emph{Physical Rev. Lett.} 71:1477--1481.

\bibitem{kardar2007statistical}
Kardar M
\newblock (2007) \emph{Statistical physics of particles}
\newblock (Cambridge Univ Pr).

\bibitem{cover1991elements}
Cover T, Thomas J, Wiley J, {et~al.}
\newblock (1991) \emph{Elements of information theory}
\newblock (Wiley Online Library) Vol.{}~6.

\bibitem{mackay2003information}
MacKay D
\newblock (2003) \emph{Information theory, inference, and learning algorithms}
\newblock (Cambridge Univ Pr).

\bibitem{touchette2009large}
Touchette H
\newblock (2009) The large deviation approach to statistical mechanics.
\newblock \emph{Physics Reports} 478:1--69.

\bibitem{bunin2011universal}
Bunin G, D'Alessio L, Kafri Y, Polkovnikov A
\newblock (2011) Universal energy fluctuations in thermally isolated driven
  systems.
\newblock \emph{Nature Physics} 7:913--917.

\bibitem{deffner2010generalized}
Deffner S, Lutz E
\newblock (2010) Generalized clausius inequality for nonequilibrium quantum
  processes.
\newblock \emph{Phys. Rev. Lett.} 105:170402.

\bibitem{deffnerlutz2011}
Deffner S, Lutz E
\newblock (2011) Nonequilibrium entropy production for open quantum systems.
\newblock \emph{Phys. Rev. Lett.} 107:140404.

\bibitem{polkovnikov2011microscopic}
Polkovnikov A
\newblock (2010) Microscopic diagonal entropy and its connection to basic
  thermodynamic relations.
\newblock \emph{Annals of Physics} 326:486--499.

\bibitem{levine1978information}
Levine R
\newblock (1978) Information theory approach to molecular reaction dynamics.
\newblock \emph{Annual Review of Physical Chemistry} 29:59--92.

\bibitem{schlogl1980}
Schl\"ogl F
\newblock (1980) Stochastic measures in nonequilibrium thermodynamics.
\newblock \emph{Phys. Rep.} 62:267.

\bibitem{qian2001relative}
Qian H
\newblock (2001) Relative entropy: free energy associated with equilibrium
  fluctuations and nonequilibrium deviations.
\newblock \emph{Phys. Rev. E} 63:042103.

\bibitem{mehta2008nonequilibrium}
Mehta P, Andrei N
\newblock (2008) Nonequilibrium quantum impurities: From entropy production to
  information theory.
\newblock \emph{Phys. Rev. Lett.} 100:86804.

\bibitem{LL5}
Landau L, Lifshitz E
\newblock (1980) \emph{Statistical Physics Part I}
\newblock (Butterworth-Heinemann).

\bibitem{berges2004prethermalization}
Berges J, Borsanyi S, Wetterich C
\newblock (2004) Prethermalization.
\newblock \emph{Physical review letters} 93:142002.


\bibitem{gurarie1995probability} Gurarie V., Probability density, diagrammatic technique, and epsilon expansion in the theory of wave turbulence, Nuclear Physics B {\bf 441}, 569 (1995).


\bibitem{moeckel2010crossover}
Moeckel M, Kehrein S
\newblock (2010) Crossover from adiabatic to sudden interaction quenches in the
  hubbard model: prethermalization and non-equilibrium dynamics.
\newblock \emph{New Journal of Physics} 12:055016.

\bibitem{gring_12} M.~Gring, M.~Kuhnert, T.~Langen, T.~Kitagawa, B.~Rauer, M.~Schreitl, I.~Mazets, D.~A.~Smith, E.~Demler, J.~Schmiedmayer, Relaxation and Pre-thermalization in an Isolated Quantum System, Science {\bf 337}, 1318 (2012).


\bibitem{polkovnikov2011colloquium}
Polkovnikov A, Sengupta K, Silva A, Vengalattore M
\newblock (2011) Colloquium: Nonequilibrium dynamics of closed interacting
  quantum systems.
\newblock \emph{Reviews of Modern Physics} 83:863.


\bibitem{deutsch1991quantum}
Deutsch J
\newblock (1991) Quantum statistical mechanics in a closed system.
\newblock \emph{Physical Review A} 43:2046.

\bibitem{srednicki1994chaos}
Srednicki M
\newblock (1994) Chaos and quantum thermalization.
\newblock \emph{Phys. Rev. E} 50:888.

\bibitem{rigol2008thermalization}
Rigol M, Dunjko V, Olshanii M
\newblock (2008) Thermalization and its mechanism for generic isolated quantum
  systems.
\newblock \emph{Nature} 452:854--858.

\bibitem{jaynes1957information}
Jaynes E
\newblock (1957) Information theory and statistical mechanics. ii.
\newblock \emph{Phys. Rev.} 108:171.

\bibitem{LL1}
Landau L, Lifshitz E
\newblock (1980) \emph{Classical Mechanics}
\newblock (Butterworth-Heinemann).

\bibitem{polkovnikov2008heat}  A.~Polkovnikov, Microscopic expression for the heat in the diagonal basis,
 Phys. Rev. Lett. {\bf 101}, 220402 (2008).

\bibitem{gurarie2012entropy} V.~Gurarie, Large time dynamics and the generalized Gibbs ensemble,  arXiv:1209.3816.

\bibitem{rigol2007GGE}  M.~Rigol, V.~Dunjko, V.~Yurovsky, M.~Olshanii, Relaxation in a completely integrable many-body quantum system: An ab initio study of the dynamics of the highly excited states of 1d lattice hard-core bosons,  Phys. Rev. Lett. {\bf 98}, 050405 (2007).

\bibitem{arakilieb1970} H. Araki and E. H. Lieb,  Entropy Inequalities, Comm. Math. Phys. {\bf 18}, 160, (1970).



\end{thebibliography}
\end{document}